\documentclass[msc,oneside]{ubcthesis}

\institution{The University Of British Columbia}
\faculty{Irving K. Barber Faculty of Science}
\institutionaddress{Okanagan}

\degreetitle{B.Sc. Computer Science Honours}

\title{Towards Parallel Learned Sorting}
\author{Ivan Carvalho} 
\copyrightyear{2022}
\submitdate{April 2022} 
\program{Computer Science}

\usepackage{ubcostyle} 
\usepackage[frozencache,cachedir=minted-cache]{minted}
\usepackage{pgf}
\usepackage{colortbl}
\usepackage{caption}
\newenvironment{codechunk}{\captionsetup{type=listing}}{}

\begin{document}

\frontmatter                    

\maketitle                      

\begin{abstract}                

We introduce a new sorting algorithm that is the combination of ML-enhanced sorting with the  In-place Super Scalar Sample Sort (\(\text{IPS}^{4}\text{o}\)). The main contribution of our work is to achieve parallel ML-enhanced sorting, as previous algorithms were limited to sequential implementations. We introduce the In-Place Parallel Learned Sort (IPLS) algorithm and compare it extensively against other sorting approaches. IPLS combines the \(\text{IPS}^{4}\text{o}\) framework with linear models trained using the Fastest Minimum Conflict Degree algorithm to partition data. The experimental results do not crown IPLS as the fastest algorithm. However, they do show that IPLS is competitive among its peers and that using the \(\text{IPS}^{4}\text{o}\) framework is a promising approach towards parallel learned sorting.

\end{abstract}

\newpage
\phantomsection \label{tableofcontent}
\addcontentsline{toc}{chapter}{\contentsname}
\tableofcontents                
\newpage 
\phantomsection \label{listoftab}
\newpage
\phantomsection \label{listoffig}
\addcontentsline{toc}{chapter}{\listfigurename}
\listoffigures                  


\chapter{Acknowledgements}      

I would like to thank my supervisor, Dr. Ramon Lawrence, for all the support while completing this work. Thanks for motivating me to do research in the field of databases and Learned Indexes.

I would also like to thank my friends Matthew Currie, Sean Roarty, and Kathryn Lecha. Thanks for listening to incomplete parts of this thesis, often when the details were unclear and things went wrong.

\mainmatter


\chapter{Introduction}

Sorting algorithms are among the most fundamental algorithms in Computer
Science and are critical in database applications. The performance of
sorting can impact common operations such as index building, data
aggregation, and in-memory joins. Asymptotically optimal comparison-based
sorting algorithms such as Quicksort have been known since the 1960s
\cite{Hoare1962}. Despite that, researchers keep pushing the boundaries of
sorting performance by proposing algorithms that outperform Quicksort.

Two boundaries are relevant for this work. The first boundary is that
of CPU-based parallel sorting, with In-Place Parallel Super-Scalar
Sample Sort (\(\text{IPS}^{4}\text{o}\)) as the state-of-the-art in the
field \cite{Axtmann2022}. The second boundary is that of Machine Learning (ML)
enhanced sorting, which proposes training simple ML models to predict
the position of the data to sort items \cite{Kristo2020}.

Our contribution is the combination of ML-enhanced sorting with \(\text{IPS}^{4}\text{o}\) and extensive experimental evaluation. 

In this thesis, we introduce the In-Place Parallel Learned Sort
(IPLS) algorithm. IPLS builds upon the algorithmic framework from
\(\text{IPS}^{4}\text{o}\), replacing the decision tree with the linear
model introduced in \cite{Wu2021}. We are the first work to interpret
\(\text{IPS}^{4}\text{o}\) as a ML-enhanced algorithm, which has two
consequences. The first consequence is that models from the field of
Learned Indexes \cite{Kraska2018} can be used to create variants of
\(\text{IPS}^{4}\text{o}\). The second consequence is that
\(\text{IPS}^{4}\text{o}\) provides a framework to efficiently
parallelize ML-enhanced sorting algorithms, addressing the shortcomings
of earlier works that only implemented serial routines.

We evaluate the performance of IPLS against other sorting algorithms in a benchmark adapted from the Learned Sort 2.0 paper \cite{Kristo2021}. The extensive results do not crown IPLS as the fastest algorithm. However, they do show that IPLS is competitive among its peers and that using the \(\text{IPS}^{4}\text{o}\) framework is a promising approach towards parallel learned sorting.

\chapter{Background}

There is a multitude of sorting algorithms that leverage creative
techniques to enhance performance. Despite the sheer number of
algorithms, they can be grouped into categories due to their similarities.
We focus on two categorizations that are relevant to the two
algorithms we will review: comparison-based vs distribution-based; and
in-place vs non-in-place.

Comparison-based algorithms such as Quicksort rely on comparing pairs
of elements to sort the data. This class of algorithms has a lower bound
of \(\Omega (n \log n)\) comparisons to sort the data.
Distribution-based algorithms on the other hand are not subject to that
lower bound. Radix Sort has a complexity of \(\mathcal{O} (w n)\), where
\(w\) is the width of the key being sorted. However, we note that the
performance of sorting algorithms goes beyond just asymptotic
complexity. The algorithms we will review exploit the constants of the
implementations with techniques such as instruction parallelism, loop
unrolling, SIMD instructions and other clever tricks to be faster.
There is no clear winner between the two classes. We will verify that
Radix Sort does perform better than comparison-based sorts in sequential
settings. However, if a parallel setting is taken into account, we verify the opposite. The
fastest algorithm for the sequential case is distribution-based, while
the fastest algorithm for the parallel case is comparison-based.

In-place algorithms use a constant amount of memory to sort the array,
while non-in-place algorithms use an amount that depends on the size
\(n\) of the input. This affects the performance, especially for large
inputs. As \(n\) grows, allocating memory becomes more expensive hence
using a constant amount of memory becomes advantageous. Moreover,
in-place algorithms have different memory access patterns than
non-in-place algorithms. If implemented carefully, in-place algorithms
can leverage cache-friendly memory access patterns that boost
performance. We note that the two algorithms we review are
in-place sorting algorithms.

\section{In-Place Parallel Super-Scalar Sample Sort}

\(\text{IPS}^{4}\text{o}\) \cite{Axtmann2022} is an enhanced in-place and parallel
version of the Super Scalar Sample Sort algorithm \cite{Sanders2004}. Sample Sort
itself is a generalization of Quicksort to multiple pivots. It uses
\(k\) pivots instead of just one to perform the partitioning and runs
in \(\mathcal{O}(n \log {n})\).

Partitioning works in three steps. The first step is to sample
\(\alpha k - 1\) random elements from the array, where \(\alpha\) is an
oversampling factor that can be tweaked. The second step is to sort the
sampled elements and pick \(k\) equidistant keys. The third step is to
predict a bucket \(\mathcal{B}_{i}\) for each key \(x_{i}\) and move
\(x_{i}\) to the allocated region of \(\mathcal{B}_{i}\) in the array.
After every element has been moved into its bucket, each bucket is an
independent sorting sub-problem that is handled recursively.

The difference between \(\text{IPS}^{4}\text{o}\) and previous variants
of Sample Sort is that the partitioning is done in-place. The previous
versions allocated \(\mathcal{O}(n)\) memory for an auxiliary array that
stores a copy of the elements. \(\text{IPS}^{4}\text{o}\) instead uses
buffers of size \(b\) for each bucket. It allocates \(\mathcal{O}(kb)\)
total memory, and when a buffer is full it flushes back to the array.

On a first pass, \(\text{IPS}^{4}\text{o}\) predicts a bucket for each
key and creates \(\mathcal{O}(n/b)\) blocks of size \(b\). Afterwards,
\(\text{IPS}^{4}\text{o}\) permutes the blocks such that each region
\(\mathcal{B}_{i}\) is contiguous in the array using a routine similar
to defragmentation.

The benefits of the in-place partitioning routine are twofold. The
immediate benefit is that the required memory to execute the algorithm
is lower. The second benefit is the cache-friendly pattern of the memory
accesses. \(b\) can be chosen carefully such that the buffers fit into
the cache, which reduces cache misses. In contrast, the previous
versions of Sample Sort accessed scattered memory positions which is
slower.

In addition to being in-place, \(\text{IPS}^{4}\text{o}\) is also
parallelizable. The authors of \(\text{IPS}^{4}\text{o}\) implemented an
efficient task scheduler to coordinate the parallel tasks with the goal
of always trying to use most of the hardware available. Both the
prediction step and the block permutation step in the partitioning phase
can be executed in parallel using tasks. Once the recursive sub-problem
sizes become small, sorting the sub-problem sequentially also becomes a
task. The end result of parallelizing all steps of the algorithm is an
algorithm that outperforms Quicksort by a large margin in multi-threaded
settings.

Another relevant contribution from the authors of
\(\text{IPS}^{4}\text{o}\) is that they reuse the code from the in-place
partitioning and task scheduler to implement Radix Sort as well. The
In-place Parallel Super Scalar Radix Sort (\(\text{IPS}^{2}\text{Ra}\))
builds on top of the \(\text{IPS}^{4}\text{o}\) and uses the
most-significant digits to partition the data instead of using sampled
pivots. The benchmarks for the authors indicate that
\(\text{IPS}^{2}\text{Ra}\) consistently beats
\(\text{IPS}^{4}\text{o}\) in sequential settings and is among the
fastest Radix Sort implementations available.

\section{Learned Sort}

Learned Sort is a distribution-based sorting algorithm that introduces
the paradigm of ML-enhanced sorting \cite{Kristo2020}. The main idea is that if
there exists a model \(F\) that predicts the sorted position of a key
\(x\), we could sort the array in a single pass by moving each element
to its correct location with \(A[F(x)] = x\).

\begin{figure}[ht]

\begin{center}
\includegraphics[width=0.55\linewidth]{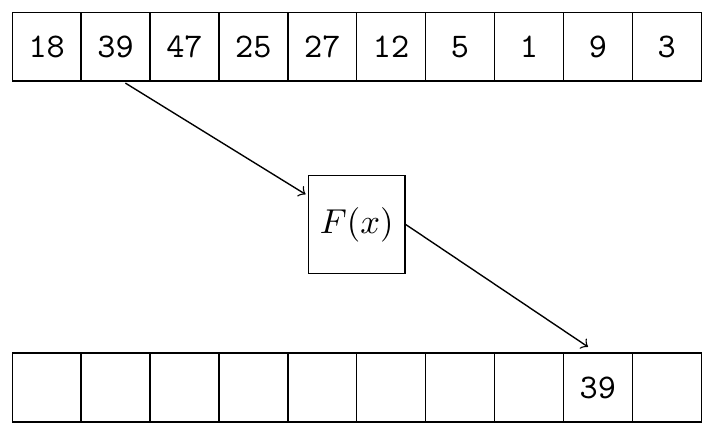} 
\end{center}

\caption{\label{fig:idealsorting} Ideal case of ML-enhanced with a perfect model.}\label{fig:unnamed-chunk-2}
\end{figure}

The first challenge with this idea is that \(F\) is unknown at the
start. Instead of using the exact model,
we sample some data from \(A\) and train a model on those samples. The
second challenge with the idea is picking the appropriate model that can
predict the sorted position of an element. One possible solution is to
use Empirical Cumulative Distribute Functions (eCDF). The eCDF yields
the probability \(P(A \leq x)\) that an element is smaller than \(x\),
hence for an array with \(N\) elements we predict that the position will
be \(pos = F(x) = \lfloor N \cdot P(A \leq x) \rfloor\).

Learned Sort relies on the Recursive Model Index (RMI) architecture
introduced in \cite{Kraska2018} to calculate the eCDF. RMIs were originally
proposed for static index lookups in in-memory databases. They were
originally trained on the sorted array containing all data to be
indexed, and performed look-ups for keys \(x\) by checking if \(x\) was
in the range \([F(x) - \varepsilon, F(x) + \varepsilon]\) of the array where
\(\varepsilon\) is the error of the RMI. Learned Sort does not have all the
data sorted in advance, as the input by definition needs to be sorted.
The algorithm overcomes this issue by sorting the sampled data using
other algorithms and training the RMI on the sorted sample. There is a
trade-off between sample size, the accuracy of the eCDF, and the
training time. If a sample size is too small, it trains faster but
yields worse estimates for the eCDF. If the training size is too big,
the eCDF estimate is better but sorting the samples dominates the
run-time. The authors of Learned Sort found that sampling 1\% worked
well in practice.

Additional details need to be handled to make ML-enhanced sorting work
in practice. The eCDF is an estimate that is not guaranteed to be
monotonic, therefore there might exist elements \(a\) and \(b\) such
that \(a < b\) but \(F(a) > F(b)\), and the implementation needs to
correct those errors. Moreover, there can be collisions with elements
having \(F(a) = F(b)\), hence the algorithm also needs to account for
those. The collision problem is exacerbated when there are many
duplicates because in that case it is guaranteed that all duplicates
will collide at \(F(x)\).

Learned Sort 2.0 \cite{Kristo2021} accomplishes the task of applying ML-enhanced
sorting in practice. The algorithm consists of three routines:
Partitioning, Model-Based Counting Sort, and Correction with Insertion
Sort. The partitioning routine is in-place and uses a strategy similar
to the partitioning introduced in \(\text{IPS}^{4}\text{o}\). The main
differences are that the eCDF is used to assign keys to buckets, that
the number of buckets \(k\) is higher compared to
\(\text{IPS}^{4}\text{o}\), and that partitioning is executed only twice
instead of recursively. To handle duplicates, Learned Sort 2.0 performs
a homogeneity check after partitioning: if all elements within a bucket
are equal, the bucket is left as is because a sequence of identical
elements is already sorted. The base case is a Model-Based Counting Sort
that uses the eCDF to predict the final position of the keys in the
buckets. Lastly, Insertion Sort is executed to correct the possible
mistakes from the eCDF and guarantee that the output is sorted. Because
the sequence is almost sorted, Insertion Sort is cheap to execute in
practice.

Learned Sort 2.0 outperforms Quicksort and is competitive with
state-of-the-art algorithms such as \(\text{IPS}^{4}\text{o}\). This
fact is remarkable, given that ML-enhanced sorting is a novel approach
competing against implementations that have been extensively tuned.
There is room for algorithmic innovation in ML-enhanced sorting, hence
future work might make it even more competitive with other approaches.

There are limitations to using Learned Sort. An obvious
limitation is that the current implementations focus on numeric types
and cannot sort strings. The most substantial limitation of
Learned Sort currently is that it does not support parallelization. This
can limit its use in practice because parallel algorithms such as
\(\text{IPS}^{4}\text{o}\) can outperform Learned Sort by using all the
hardware available.

\section{Related Work}

ML-enhanced sorting is closely related to the field of Learned Indexes \cite{Kraska2018}. Learned Indexes come from an emerging field using machine learning to create highly efficient index data structures that outperform traditional data structures such as B-Trees. 

Early work on learned indexes focused mainly on key lookup and primarily did not support updates. Their models relied heavily on modeling the eCDF to predict the correct location of a key on a sorted array, albeit with different strategies. RMIs count on a two-layer model where each model at the second layer can be seen as an expert on a specific chunk of the eCDF. RadixSpline \cite{Kipf2020} approximates the eCDF using piecewise linear functions and a radix table. The linear functions of RadixSpline guarantee a maximum error of \(\varepsilon\) for the eCDF approximation, hence to find a key we use the radix prefix of the element to use the correct function to approximate the eCDF. The Piecewise Geometric Model Index \cite{Ferragina2020} also uses the idea of piecewise linear functions with bounded error to approximate the eCDF, although it opts for a multi-level structure instead of a radix table.

Later work on learned indexes also incorporated indexes that supported updates such as ALEX \cite{Ding2020} and LIPP \cite{Wu2021}. Both ALEX and LIPP store the data in a tree structure and use models to navigate the tree. Due to the tree structure, those indexes are also concerned with keeping a small tree depth. To achieve small depths, they aim to minimize the maximum number of elements in a tree node. Hence, newer models from learned indexes also try to achieve a balanced partitioning. 

Other works that are related to this thesis are the Learned In-Memory Joins \cite{Sabek2021}, particularly Learned Sort-Join. Learned Sort-Join profits from the eCDF modeling approach discussed earlier to speed up both the partitioning and the chunked-join phases of the algorithm.

\chapter{Machine Learning Models for Sorting}

ML-enhanced sorting algorithms leverage statistical properties of the distribution to be sorted to achieve faster execution times. In general, ML-enhanced sorting is characterized by four phases:

\[
\text{Sample Data} \rightarrow \text{Sort Sample} \rightarrow \text{Train Model} \rightarrow \text{Predict on Keys}
\]

Before we review each phase, it is important to introduce the concept of computing budget for ML-enhanced sorting. ML-enhanced sorting competes against traditional sorting algorithms like Quicksort. Hence, the cost of executing all four phases from ML-enhanced sorting must be less than or equal to the cost of executing Quicksort. This is a narrow budget because Quicksort itself is an efficient algorithm. The consequence of the computing budget is that some choices that in theory lead to better models need to be ruled out because they would cause the algorithm to be slower than Quicksort. 

The first phase is sampling data. This phase is required because every model needs data to train on. The size of the sample is highly dependent on the algorithm and the model used. More samples lead to more accurate models, however, there is a cost to sampling more data as we will explain in the next phase.

The second phase is sorting the sample using another sorting algorithm. The models used in sorting rely on the relative order of the elements from the sample to learn about the data distribution. Hence, the sample needs to be sorted which is often done using a third algorithm. The cost of the second phase will always be higher than the cost of the first phase. For \(S\) samples, the first phase runs in \(\mathcal{O}(S)\) time while the second phase will generally run in \(\mathcal{O}(S \log S)\).

The second phase is also relevant for implementation purposes because it implies that ML-enhanced algorithms cannot exist on their own. ML-enhanced algorithms can recursively call themselves to sort the sample if the sample size is large. However, a traditional sorting algorithm will still be at the base. The fact is not necessarily an issue, because many popular sorting implementations use a third algorithm such as Insertion Sort to handle base cases.

The third phase is training the model. This phase takes at least \(\Omega (S)\) work, but its cost is very dependent on the chosen model. ML-enhanced sorting favors models that can be trained quickly and that do not require lots of parameters. Often, there is no budget to do a complete search to find the optimal parameters for a model hence the search space is reduced. An even more radical approach is to set some parameters to constants that work well in the general case, such as in Learned Sort.

Moreover, we can classify the models we can train into two types. Partitioning models are trained to predict the bucket \(\mathcal{B}_{i}\) of an element. The output of prediction-based models is an integer in \(\{0, 1, 2, 	\ldots k - 1 \} \) where \(k\) is the number of buckets. This type of model aims to minimize metrics related to the number of elements per bucket in order to achieve a balanced partitioning. A first metric is to minimize the average number of elements per bucket, \(\sum_{i=0}^{k-1} (|\mathcal{B}_{i}| - \frac{N}{k})^{2}\). Another possible metric is to minimize the maximum number of elements in a bucket, \(\min ( \max_{0 \leq i < k} ( |\mathcal{B}_{i}| ) ) \). We point out that the choice of the metric impacts the cost of training the model: for example, it is cheaper to use the second metric to train linear models.

In contrast, eCDF models are trained to predict \(P(A \leq x)\), the proportion of elements that are smaller than a given element in the distribution. This happens because with \(P(A \leq x)\), we can scale the output of the eCDF to find the exact position of \(x\) in the array. The outputs of this type of model are real numbers  in \([0, 1)\). eCDF models generally aim to minimize the mean-squared error of their predictions, \( \frac{1}{N} \sum_{i=0}^{N-1} (F(x_{i}) - P(A \leq x_{i}))^{2}\). We highlight that any eCDF model can be converted to a partitioning model of \(k\) buckets by scaling using \(\mathcal{B}_{i} = \lfloor k \cdot F(x_{i})\rfloor\).

The last phase is predicting the position or bucket of a key using the trained model. For partition-based models, this is the phase that effectively creates independent sorting subproblems that can be handled recursively. For eCDF-based models, this phase is the one that attempts to sort the array by putting the elements in their correct positions.

The cost of the predicting phase is again dependent on the chosen model, just like in the training phase. However, we need to take into account that the training phase handles \(S\) samples while the predicting phase handles all the \(N\) elements of the array. In that sense, slow predicting times are amplified by the fact that they apply to many elements. ML-enhanced sorting favors again simpler models such as those with \(\mathcal{O}(1)\) predicting time per element.

\section{Recursive Model Index}

The Recursive Model Index (RMI) is an architecture based on a hierarchy of
models that predicts the eCDF. It emerged as an alternative for B-Trees
when doing look-ups on in-memory static arrays. Because RMIs estimate
the eCDF, they can also be used for sorting in algorithms that rely on
eCDF such as Learned Sort.

A RMI consists of \(L\) levels, and each level \(i\) has \(M_{i}\)
models that recursively select the model in the next level of the RMI.
The output \(F(x)\) of an RMI is a value in \([0, 1)\) that is an
approximation for \(P(A \leq x)\).

\begin{figure}[ht]
\begin{center}
\includegraphics[width=0.67\textwidth]{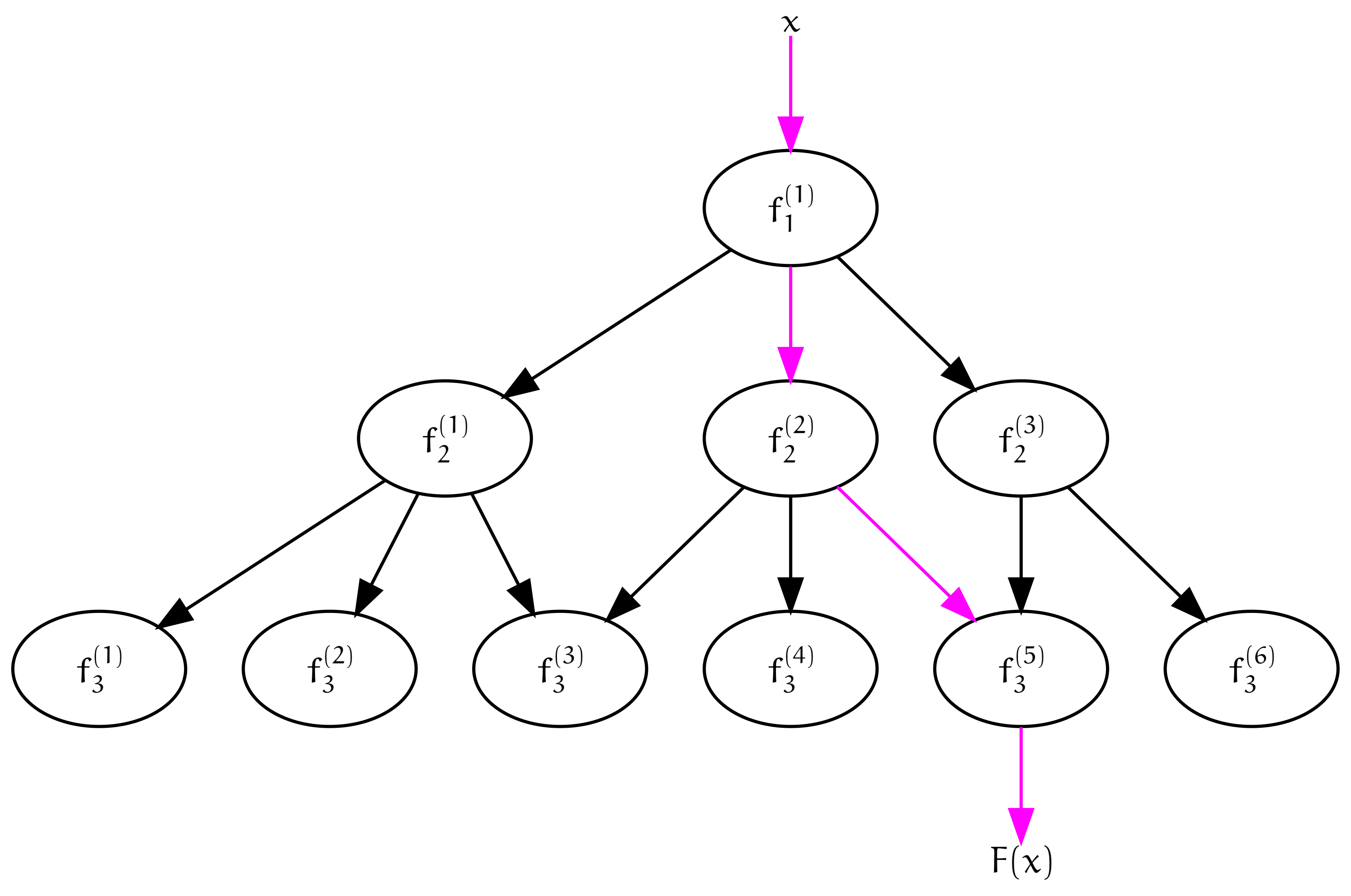}
\end{center}
\caption{\label{fig:rmi1} Generic RMI with multiple levels.}\label{fig:unnamed-chunk-4}
\end{figure}

We denote the \(k\text{-th}\) model of level \(i\) as \(f_{i}^{(k)}(x)\)
and the ensemble of the level as \(f_{i}(x)\) with
\(f_{i}(x) \in [0, 1)\). Hence, \(i\text{-th}\) level of a RMI can be
described mathematically using recursion:

\[
f_{i}(x) = f_{i}^{(\lfloor M_{i} f_{i - 1}(x) \rfloor )} (x) 
\]

The base case is at the first level with \(f_{1}(x) = f_{1}^{(1)}(x)\),
and the eCDF is given by \(F(x) = f_{L}(x)\). The recursive strategy of
the RMI is particularly successful at avoiding the Last Mile Problem.
The Last Mile Problem is the challenge to reduce the magnitude of the
error of the eCDF approximation. For example, reducing the error rate
from \(10^{-3}\) to \(10^{-4}\) might require a much more complex model
than reducing the error rate from \(10^{-1}\) to \(10^{-2}\). RMIs
mitigate the effect of the last mile problem by ``splitting'' the data
among the models. Each model can be seen as an expert on a specific part
of the eCDF, which helps produce reasonable eCDF estimates with simpler
models.

One observation from the definition of RMIs is that the architecture is
extremely flexible. This is advantageous because, in theory, RMIs can
model the eCDF accurately. The disadvantage is that training an optimal
RMI can be a challenge. There are many parameters to optimize such as
the number of levels, the number of models for each level, and the types
of models for each \(f_{i}^{(k)}\).

In practice, RMIs used in Learned Indexes and Sorting are simpler than
the originally proposed idea and are fixed to \(L = 2\) levels:
\(F(x) = f_{2}^{(\lfloor M_{2} f_{1}^{(1)}(x) \rfloor)}(x)\).

\begin{figure}[ht]

\begin{center}
\includegraphics[width=0.60\textwidth]{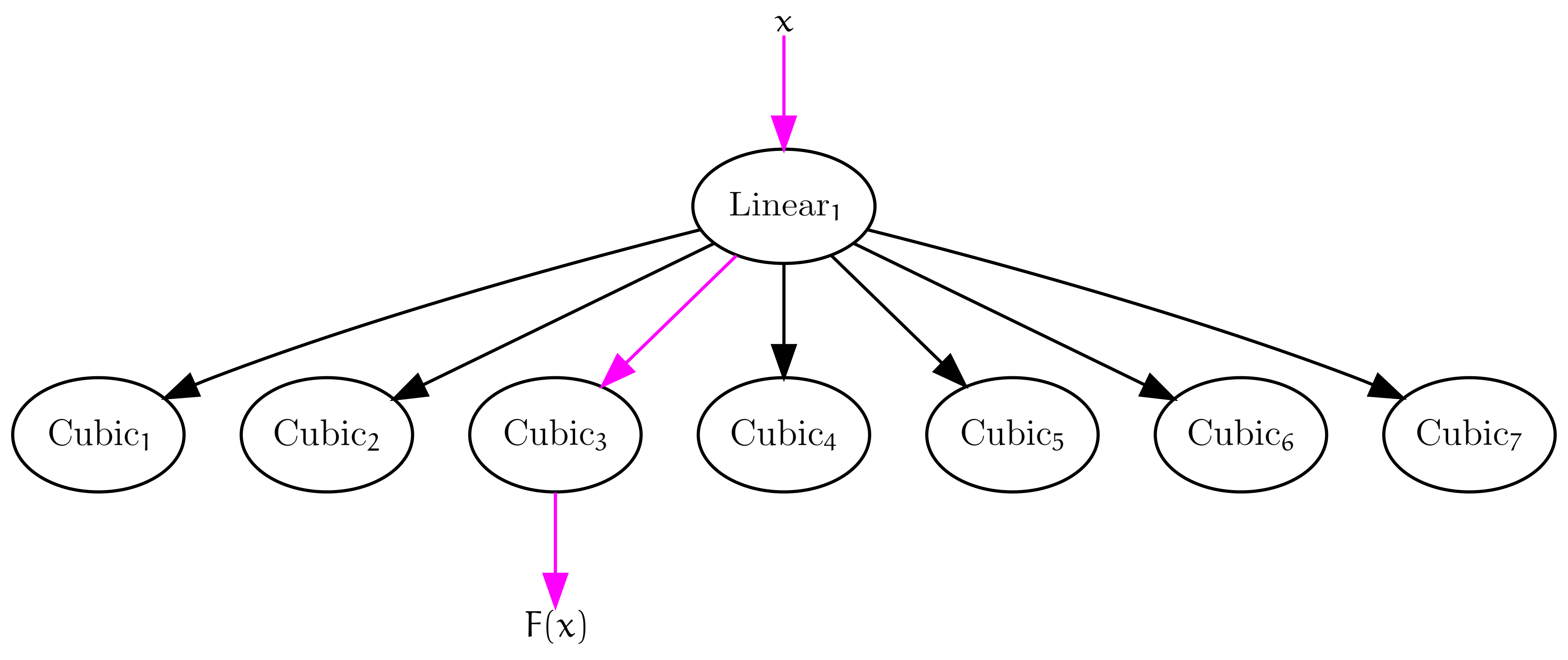}
\end{center}

\caption{\label{fig:rmi2} RMI used in practice. The number of levels is fixed to two and the optimal model types are chosen via training with heuristics.}\label{fig:unnamed-chunk-5}
\end{figure}

Despite having a fixed number of levels, finding the optimal RMI remains
a non-trivial task. We still need to choose the model type of the first
layer, the second layer, and the number of models. Because the parameters
are not independent, heuristics such as the ones introduced in CDFSHOP
\cite{Marcus2020_CDF} are needed to find the best RMI.

ML-enhanced sorting cannot afford the cost to find the optimal RMI,
because optimizing the RMI would cost more than sorting the array with
other algorithms. Fortunately, an optimal RMI is not needed to achieve
good results. The authors of Learned Sort use a linear spline on the
first layer with a fixed size of linear functions on the second layer.
The second layer functions are not trained using linear regression and
instead use linear interpolation to find the coefficients. These choices
yield a sub-optimal RMI, but the model is still effective because the
error in eCDF predictions is balanced by the quicker training time of a
constrained RMI. An independent review of the RMI found that some data
distributions may lead to many empty segments in the second layer
\cite{Maltry2021}, which could impact the efficiency of sorting because
empty segments waste the predictive power of the RMI. Nevertheless, RMIs
work well in practice and Learned Sort has competitive results.

\section{Decision Trees}

Decision Trees are non-parametric machine learning models that use comparisons
to make predictions. They are built on top of \(\mathcal{O}(k)\) training elements and have a height of \(\mathcal{O}(\log k)\). For Sorting applications, Decision Trees predict the partition bucket \(\mathcal{B}_{i}\) for each \(x_{i}\) with the property that if \(x_{i} < x_{j}\), then \(\mathcal{B}_{i} \leq \mathcal{B}_{j}\).

Training a Decision Tree is straightforward using a divide-and-conquer strategy. We can assume that the sampled data is sorted from the previous steps, which just requires converting the sorted array to a Binary Tree. At each step, we select the middle element as the splitter for that node of the decision tree and continue to build the tree recursively by dividing the array in half.

Naive implementations of the decision tree use conditionals to predict the bucket of an element, which can cause branch mispredictions. Optimized implementations such as the one from \(\text{IPS}^{4}\text{o}\) use a branchless version of the decision tree combined with loop unrolling. The splitters for the tree are stored in an array using a binary tree structure, hence the child nodes of \(a_{j}\) are \(a_{2j}\) and \(a_{2j + 1}\). Thus, the next node can be decided using just comparisons with \(j = 2j + (x > a_{j}) \). The branchless optimization is relevant because it can be combined with loop unrolling, effectively turning the prediction cost \(\mathcal{O}(1)\) for small \(k\).

Decision trees can also benefit from instruction-level parallelism. \(\text{IPS}^{4}\text{o}\) batches the elements in groups of 8 and applies loop unrolling again when predicting for the batch. This can increase the  throughput of the decision tree because modern CPUs can execute more than one instruction per clock cycle. 

One advantage of decision trees is that they are robust models against distributions with lots of duplicates. \(\text{IPS}^{4}\text{o}\) leverages the concept of equality buckets in the decision tree. If during sampling we notice that there are many duplicates of an element \(s_{i}\), we can create a bucket \(\mathcal{B}_{i}\) just for the elements \(x = s_{i}\). This is extremely convenient because after the partitioning the equality bucket \(\mathcal{B}_{i}\) will already be sorted as all the elements on it are identical. Decision trees, therefore, contrast with the other models that often struggle to handle duplicates.

The disadvantages of decision trees come from the same points that make it a straightforward, robust model. To enhance a decision tree, we need to increase \(k\) and the sample size as there are no other parameters to tweak. However, doing so increases both the training time and the evaluation time. In addition, large values of \(k\) may lead to the splitters not fitting into the cache, which can harm performance. Thus, decision trees are efficient when the number of buckets is moderate, but inefficient if the number of buckets is high.

\section{Linear Models}

Linear models are among the most fundamental machine learning models and they can also be applied to sorting. Evaluating a linear model is trivial and can be done in \(\mathcal{O}(1)\), which is a desirable property for sorting as the computing budget for prediction time is low. Another desirable property of linear models is monotonicity, because if \(x_{i} < x_{j}\), then \(F(x_{i}) \leq F(x_{j})\) (or the mirrored equivalent if the slope is negative).

We focus on linear models for partitioning because linear models for eCDF are dominated by RMIs. Each RMI can potentially have a linear model at \(f_{i}^{(k)}(x)\) and it is very likely that a two-layer RMI performs better than a single linear model.

The models for partitioning are described by three parameters: the number of buckets \(k\), the slope \(a\) and the constant term \(b\). Mathematically, the model is:

\[
F(x) = \begin{cases}
    0, & \text{if} \hspace{0.2cm} \lfloor a \cdot x + b \rfloor < 0\\
    k - 1, & \text{if} \hspace{0.2cm} \lfloor a \cdot x + b \rfloor \geq k\\
    \lfloor a \cdot x + b \rfloor, & \text{otherwise}
\end{cases}
\]

The equation is adjusted using the floor function because the output must always be an integer. The result needs to satisfy \(0 \leq a \cdot x + b < k\), hence if the output is negative or too large we map it to zero and \(k - 1\) respectively.

Some choices must be made when training a linear model because there are multiple metrics to optimize when training a linear model. Ideally, we aim to have \(\frac{N}{k}\) elements per bucket if the goal is to have balanced partitioning. Hence, one possible metric is minimizing \(\sum_{i=0}^{k-1} (|\mathcal{B}_{i}| - \frac{N}{k})^{2}\). Minimizing that specific metric is a challenge, hence we analyse two alternatives.

The first option is minimizing the mean-squared error of the predicted bucket using a traditional linear regression. This is a valid option and the regression can be trained in \(\mathcal{O}(S)\) for \(S\) samples. If we minimize the sum of the squared distances between each element and its target bucket, we expect that most elements will indeed fall into their bucket. However, this strategy may lead to a bucket having too many elements. Having too many elements on a bucket implies doing more recursive partition steps. 

The second option is to minimize the maximum number of nodes in a bucket. The Fastest Minimum Conflict Degree (FMCD) is an algorithm introduced in \cite{Wu2021} that trains a linear model with exactly that property. It was originally proposed as the model for the Updatable Learned Index with Precise Positions (LIPP), a novel learned index that outperforms RMIs. FMCD can train a linear model from \(S\) sample data points in \(\mathcal{O}(S)\) time and guarantees that at most \(\frac{S}{3}\) elements will be in the same bucket if there are no duplicate data points. If the sample is representative, we also expect that at most \(\frac{N}{3}\) keys will be in the bucket with the most elements after applying the model to every element.

\begin{codechunk}
\begin{minted}{c++} 
template<typename Iterator>
std::tuple<double, double, std::size_t> FMCD(
    Iterator begin, Iterator end, std::size_t K
) {
    std::size_t i = 0;
    std::size_t D = 1; 
    std::size_t N = static_cast<std::size_t>(end - begin);

    long double u_d = (
        static_cast<long double>(begin[N - 1 - D])
        - static_cast<long double>(begin[D])
    )/static_cast<long double>(K - 2);

    while(i <= N - 1 - D){
        
        while(
            i + D < N 
            && static_cast<long double>(begin[i + D] - begin[i]) >= u_d
        ){
            i++;
        }

        if(i + D >= N) { break; }

        D += 1;
        if(D*3 > N) {
            break;
        }

        u_d = (
            static_cast<long double>(
            begin[N - 1 - D]) 
            - static_cast<long double>(begin[D])
        )/static_cast<long double>(K - 2);

    }

    long double A = 1/u_d;
    long double B = (K - A * (
        static_cast<long double>(
        begin[N - 1 - D]) 
        + static_cast<long double>(begin[D])
    )) / static_cast<long double>(2);

    return std::make_tuple(
        static_cast<double>(A), static_cast<double>(B), D
    );

}
\end{minted} 
\captionof{listing}{Implementation of FMCD in C++ \linebreak}
\end{codechunk}

The linear model trained by FMCD is remarkable because it is suitable for sorting even when the input distribution is non-linear. The \(\frac{N}{3}\) bound yields that on average \(\mathcal{O}(\log N)\) recursive partition steps will be performed no matter what input is given. If the input is modeled by a linear model, then it is likely that fewer steps will be required because the constant of the logarithm will be smaller.

In addition, we highlight that using large values for \(k\) is feasible using linear models. The prediction time is independent of \(k\), therefore the cost for increasing \(k\) only affects the training step. This makes linear models good candidates for parallel sorting on a large number of cores, as parallel sorting can benefit from a large \(k\) to match the number of cores \(p\).

\chapter{Proposed Algorithm}

We propose the In-Place Parallel Learned Sort (IPLS) algorithm. IPLS extends  \(\text{IPS}^{4}\text{o}\) and has the goal to show that the code from \(\text{IPS}^{4}\text{o}\) can be used to implement a parallel version of ML-enhanced sorting. IPLS keeps most of the details of in-place partitioning and parallelism from \(\text{IPS}^{4}\text{o}\), but changes the model and the base case for the algorithm.

IPLS partitions the data in \(k = 256\) buckets using linear models. It samples \(\alpha k - 1\) random elements from the array with \(\alpha = 0.2 \log N\) (\(k\) and \(\alpha\) were kept the same as in \(\text{IPS}^{4}\text{o}\)). Then, it trains a linear model on the samples using the FMCD algorithm discussed earlier. The linear model \(F(x)\) is then used to predict the bucket for each element. 

For \(n \leq 2^{12}\), IPLS uses SkaSort \cite{skasort2016} as the base case. SkaSort is one of the fastest Radix Sort implementations for inputs of that size. The choice of using SkaSort was inspired by \(\text{IPS}^{2}\text{Ra}\), which also uses SkaSort as a base case.

\section{Worked Example of Proposed Algorithm}

To illustrate the proposed algorithm, we give an example on a small input of size \(n = 20\) . We tweak some parameters because the algorithm would generally not be executed for a small input. Assume that there are \(k = 4\) buckets, that  the buffer size for flushing the bucket to the array is \(2\) and that the base case is executed for inputs smaller than 10.

Initially, we have an unordered array:

\[
\resizebox{\textwidth}{!}{
\begin{tabular}{|c|c|c|c|c|c|c|c|c|c|c|c|c|c|c|c|c|c|c|c|}
    \hline
    18 & 39 & 33 & 28 & 25 & 34 & 11 & 27 & 47 & 2 & 48 & 50 & 10 & 6 & 36 & 13 & 9 & 12 & 22 & 29 \\
    \hline
\end{tabular}}
\]

We sample some data and obtain the coefficients \(a = 0.15\) and \(b = -1.55\) for our linear model. The algorithm then predicts the bucket of each key. For example, for the key 28, we predict that it is in bucket 2 because \( \lfloor 0.15 \cdot 28 - 1.55 \rfloor = \lfloor 2.65 \rfloor = 2 \).

We visually represent each bucket with colors: the zeroth bucket is represented in pink, the first in yellow, the second one in green, and the third one in cyan. We recall that the buckets are zero-indexed.

\[
\resizebox{\textwidth}{!}{
\begin{tabular}{|>{\columncolor{cyan!}}c|>{\columncolor{cyan!}}c|>{\columncolor{green!}}c|>{\columncolor{green!}}c|>{\columncolor{cyan!}}c|>{\columncolor{cyan!}}c|>{\columncolor{pink!}}c|>{\columncolor{pink!}}c|>{\columncolor{cyan!}}c|>{\columncolor{cyan!}}c|>{\columncolor{pink!}}c|>{\columncolor{pink!}}c|>{\columncolor{pink!}}c|>{\columncolor{pink!}}c|>{\columncolor{yellow!}}c|>{\columncolor{yellow!}}c|>{\columncolor{green!}}c|>{\columncolor{green!}}c|>{\columncolor{pink!}}c|>{\columncolor{cyan!}}c|}
    \hline
    39 & 33 & 28 & 25 & 34 & 47 & 11 & 2 & 48 & 50 & 10 & 6 & 13 & 9 & 18 & 22 & 27 & 29 & 12 & 36 \\
    \hline
\end{tabular}}
\]

After, we apply a pass to make the buckets contiguous:

\[
\resizebox{\textwidth}{!}{
\begin{tabular}{|>{\columncolor{pink!}}c
|>{\columncolor{pink!}}c
|>{\columncolor{pink!}}c
|>{\columncolor{pink!}}c
|>{\columncolor{pink!}}c
|>{\columncolor{pink!}}c
|>{\columncolor{pink!}}c
|>{\columncolor{yellow!}}c
|>{\columncolor{yellow!}}c
|>{\columncolor{green!}}c
|>{\columncolor{green!}}c
|>{\columncolor{green!}}c
|>{\columncolor{green!}}c
|>{\columncolor{cyan!}}c
|>{\columncolor{cyan!}}c
|>{\columncolor{cyan!}}c
|>{\columncolor{cyan!}}c
|>{\columncolor{cyan!}}c
|>{\columncolor{cyan!}}c
|>{\columncolor{cyan!}}c|}
    \hline
    11 & 2 & 10 & 6 & 13 & 9 & 12 & 18 & 22 & 28 & 25 & 27 & 29 & 39 & 33 & 34 & 47 & 48 & 50 & 36 \\
    \hline
\end{tabular}}
\]

The contiguous buckets may now be considered independent subproblems. The algorithm then opts to sort each bucket individually using the base case given that the inputs are tiny:

\[
\resizebox{\textwidth}{!}{
\begin{tabular}{|>{\columncolor{pink!}}c
|>{\columncolor{pink!}}c
|>{\columncolor{pink!}}c
|>{\columncolor{pink!}}c
|>{\columncolor{pink!}}c
|>{\columncolor{pink!}}c
|>{\columncolor{pink!}}c
|>{\columncolor{yellow!}}c
|>{\columncolor{yellow!}}c
|>{\columncolor{green!}}c
|>{\columncolor{green!}}c
|>{\columncolor{green!}}c
|>{\columncolor{green!}}c
|>{\columncolor{cyan!}}c
|>{\columncolor{cyan!}}c
|>{\columncolor{cyan!}}c
|>{\columncolor{cyan!}}c
|>{\columncolor{cyan!}}c
|>{\columncolor{cyan!}}c
|>{\columncolor{cyan!}}c|}
    \hline
    2 & 6 & 9 & 10 & 11 & 12 & 13 & 18 & 22 & 25 & 27 & 28 & 29 & 33 & 34 & 36 & 39 & 47 & 48 & 50 \\
    \hline
\end{tabular}}
\]

The algorithm terminates after executing the base case. The output is an ordered array.

\chapter{Experimental Evaluation}

We evaluate the performance of IPLS compared to other sorting algorithms
on a similar benchmark presented in the Learned Sort 2.0 paper \cite{Kristo2021}.
We executed the benchmarks on the \textbf{m5zn.metal} instance from AWS
for reproducibility. The instance runs an Intel® Xeon® Platinum 8252C CPU
@ 3.80GHz with 48 cores, 768KB of L1 cache, 24MB of L2 cache, and 192 GB
of RAM.

We compare IPLS against four algorithms: \(\text{IPS}^{4}\text{o}\), \(\text{IPS}^{2}\text{Ra}\), Learned Sort, and \texttt{std::sort} from C++ STL \cite{ISO2020}. The implementations were written in C++ and compiled with GCC 11 with the -O3 flag.

The benchmark includes sequential and parallel settings. We refer to the sequential versions of the algorithms  as ILS, \(\text{IS}^{4}\text{o}\), and \(\text{IS}^{2}\text{Ra}\). We also drop Learned Sort from the parallel benchmark because there is no parallel implementation of Learned Sort. 

The datasets used in the benchmark are split into real-world data and synthetic data. The real-world datasets contain 64-bit unsigned integer elements to be sorted. The synthetic datasets contain 64-bit double floating-point elements. We note that \(\text{IPS}^{2}\text{Ra}\) used the extractor from SkaSort that maps double to integers in order to sort the synthetic datasets.

An overview of the datasets is:

\paragraph{Real-World Datasets}

\begin{itemize}
\item
  \textbf{OSM/Cell\_IDs (N = \(2 \cdot 10^{8}\))}: Uniformly sampled
  location IDs from OpenStreetMaps \cite{Marcus2020_Benchmark}.
\item
  \textbf{Wiki/Edit (N = \(2 \cdot 10^{8}\))}: The edit timestamps from
  Wikipedia articles \cite{Marcus2020_Benchmark}.
\item
  \textbf{FB/IDs (N = \(2 \cdot 10^{8}\))}: The IDs from Facebook users
  sampled in a random walk of the network graph \cite{Marcus2020_Benchmark}.
\item
  \textbf{Books/Sales (N = \(2 \cdot 10^{8}\))}: Book popularity data
  from Amazon \cite{Marcus2020_Benchmark}.
\item
  \textbf{NYC/Pickup (N = \(10^{8}\))} : The yellow taxi trip pick-up
  time stamps \cite{Kristo2021}.
\end{itemize}

\paragraph{Synthetic Datasets}

\begin{itemize}
\item
  \textbf{Uniform (N = \(10^{8}\))}: Generated by
  \texttt{std::uniform\_real\_distribution} \cite{ISO2020} with \(a=0\) and
  \(b=N\).
\item
  \textbf{Normal (N = \(10^{8}\))}: Generated by
  \texttt{std::normal\_distribution} \cite{ISO2020} with \(\mu=0\) and
  \(\sigma=1\).
\item
  \textbf{Log-Normal (N = \(10^{8}\))}: Generated by
  \texttt{std::lognormal\_distribution} \cite{ISO2020} with \(\mu=0\) and
  \(\sigma=0.5\).
\item
  \textbf{Mix Gauss (N = \(10^{8}\))}: Generated by a random additive
  distribution of five Gaussian distributions \cite{Kristo2021}.
\item
  \textbf{Exponential (N = \(10^{8}\))}: Generated by
  \texttt{std::exponential\_distribution} \cite{ISO2020} with \(\lambda = 2\).
\item
  \textbf{Chi-Squared (N = \(10^{8}\))}: Generated by
  \texttt{std::chi\_squared\_distribution} \cite{ISO2020} with \(k = 4\).
\item
  \textbf{Root Dups (N = \(10^{8}\))}: Generated by
  \(A[i] = i \mod \sqrt{N}\) as proposed in \cite{Edelkamp2016}.
\item
  \textbf{Two Dups (N = \(10^{8}\))}: Generated by
  \(A[i] = i^2 + N / 2 \mod N\) as proposed in \cite{Edelkamp2016}.
\item
  \textbf{Zipf (N = \(10^{8}\))}: Generated by a Zipfian distribution
  with \(s_{\text{zipf}} = 0.75\) \cite{Kristo2021}.
\end{itemize}

\section{Sequential Results}

We analyse the sorting rate of the algorithms. The sorting rate is measured on keys per second and indicates the throughput of each algorithm. A higher sorting rate indicates that an algorithm is more performant.

\begin{figure}[ht!]
\[
\resizebox{\textwidth}{!}{\input{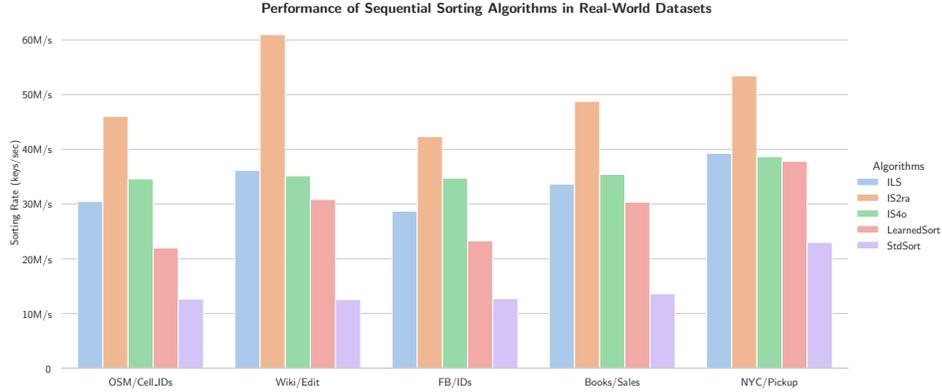}}
\]
\caption{Comparison of the sorting rate of the algorithms for real-world datasets under a sequential setting. \(\text{IS}^{2}\text{Ra}\) takes the lead in this scenario. Higher rates are better.}
\end{figure}

\(\text{IS}^{2}\text{Ra}\) is the most efficient for sorting real-world datasets. \(\text{IS}^{4}\text{o}\) comes in second and ILS comes in third place. Learned Sort comes fourth and \texttt{std::sort} comes last. Our results disagree with those from \cite{Kristo2021} that claim Learned Sort is faster than  \(\text{IS}^{4}\text{o}\) in real-world datasets. We also note that ILS beats Learned Sort in all of the real-world datasets.

\begin{figure}[ht!]
\[
\resizebox{\textwidth}{!}{\input{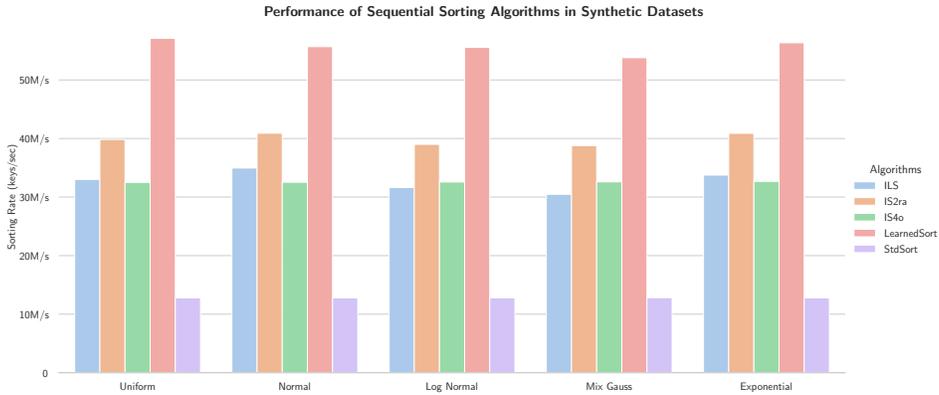}}
\]
\caption{Comparison of the sorting rate of the algorithms for synthetic datasets under a sequential setting. Learned Sort takes the lead in this scenario. Higher rates are better.}
\end{figure}

Learned Sort, however, takes the lead on synthetic datasets. \(\text{IS}^{2}\text{Ra}\) comes second. ILS and \(\text{IS}^{4}\text{o}\) tie for third place. \texttt{std::sort} comes last one more time. We can interpret the results as RMIs being able to model the distributions due to their flexibility. Compared to real-world data, distributions like the normal and exponential distribution contain less noise and are easier to model. Hence, Learned Sort achieves a favorable case which puts it first in performance. ILS does not get the same performance boost on synthetic data and presents similar sorting rates that are comparable to the ones when sorting real-world data.

\begin{figure}[ht!]
\[
\resizebox{\textwidth}{!}{\input{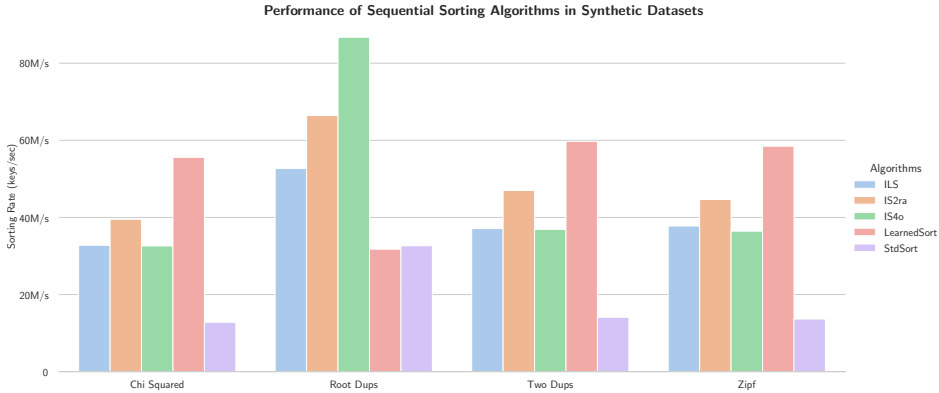}}
\]
\caption{Comparison of the sorting rate of the algorithms for synthetic with an increased number of duplicates under a sequential setting. Learned Sort takes the lead in this scenario, but struggles with the Root Dups dataset. Higher rates are better.}
\end{figure}

The trend continues and Learned Sort leads in 4 out of the 5 datasets with duplicates. However, Learned Sort struggles with the Root Dups dataset being beaten even by \texttt{std::sort}. \(\text{IS}^{4}\text{o}\) by contrast performs best on the Root Dups dataset because of the equality buckets of its decision tree. ILS remains tied with \(\text{IS}^{4}\text{o}\) in all datasets except for Root Dups because of the optimizations \(\text{IS}^{4}\text{o}\) has to handle duplicates.

\section{Parallel Results}

In the parallel setting, the fastest algorithms differ from the sequential setting. \(\text{IPS}^{4}\text{o}\) is the fastest, with IPLS in second, \(\text{IPS}^{2}\text{Ra}\) in third, and the parallel version of \texttt{std::sort} in last. In contrast to the sequential benchmark, that order was kept the same for both real-world and synthetic datasets.

\begin{figure}[H]
\[
\resizebox{\textwidth}{!}{\input{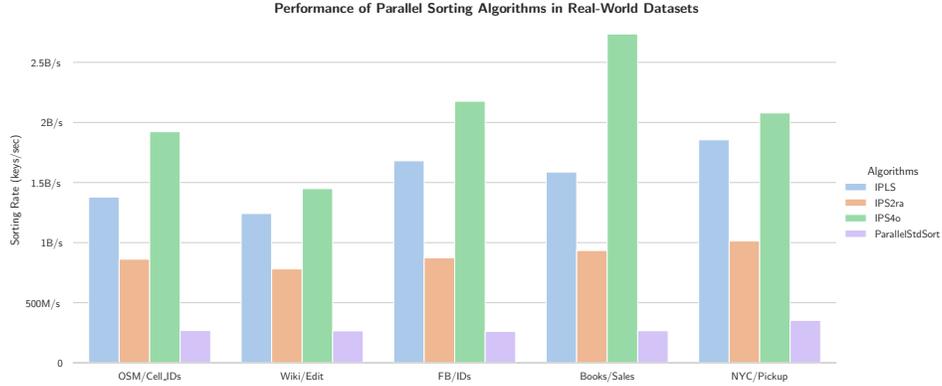}}
\]
\caption{Comparison of the sorting rate of the algorithms for real-world datasets under a parallel setting. \(\text{IPS}^{4}\text{o}\) takes the lead in this scenario. Higher rates are better.}
\end{figure}

\begin{figure}[H]
\[
\resizebox{\textwidth}{!}{\input{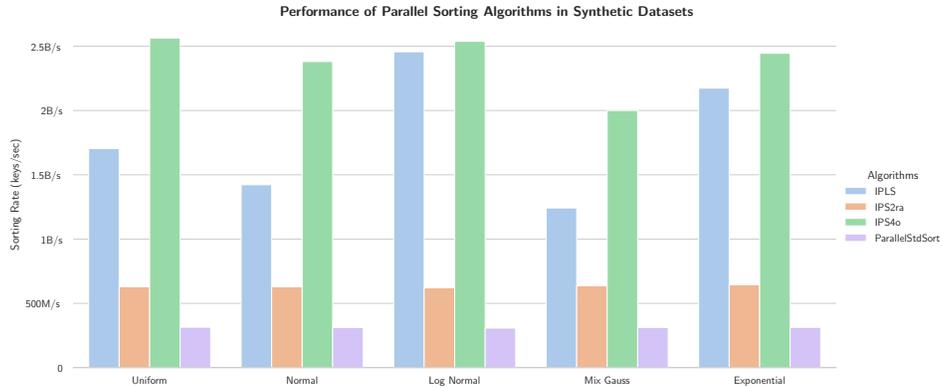}}
\]
\caption{Comparison of the sorting rate of the algorithms for synthetic datasets under a parallel setting. \(\text{IPS}^{4}\text{o}\) takes the lead in this scenario. Higher rates are better.}
\end{figure}

We may interpret the results from the parallel benchmark as to which algorithm uses the hardware available the most. \(\text{IPS}^{4}\text{o}\) uses a decision tree to partition the data, which creates many subproblems of a balanced size. This favours the performance of \(\text{IPS}^{4}\text{o}\) because it manages to keep every thread of the CPU busy always doing work.

IPLS on the other hand uses a linear model to partition the data. The linear model also creates subproblems, but some distributions might trigger the worst case of the partitioning where there is a bucket with \(N/3\) of the elements. In this sense, IPLS generally manages to use all the threads available but in some cases, it might take more recursive steps until it uses all of them.

By contrast, \(\text{IPS}^{2}\text{Ra}\) does not manage all the hardware because its partitions are not balanced. There are no bounds on the number of elements that have the same radix prefix and go in the same bucket. Hence, \(\text{IPS}^{2}\text{Ra}\) may end with threads waiting for work, hurting its sorting rate compared to \(\text{IPS}^{4}\text{o}\) and IPLS which always keep threads busy.

\begin{figure}[H]
\[
\resizebox{\textwidth}{!}{\input{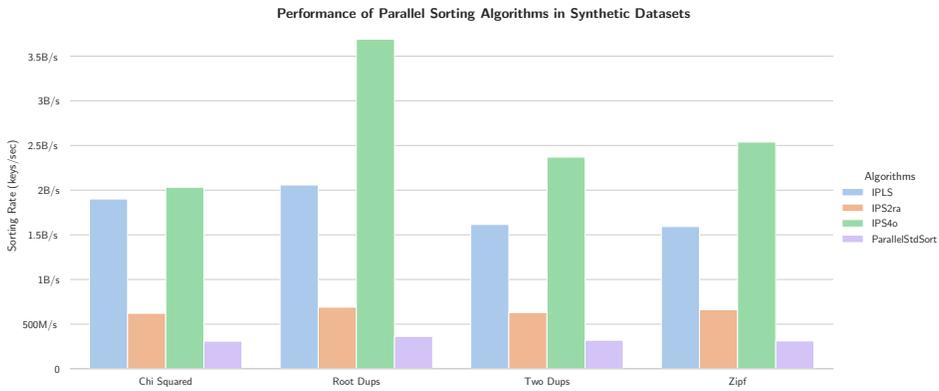}}
\]
\caption{Comparison of the sorting rate of the algorithms for synthetic datasets with an increased number of duplicates under a parallel setting. \(\text{IPS}^{4}\text{o}\) takes the lead in this scenario. Higher rates are better.}
\end{figure}

\section{Scalability}

We also compare how the sorting of the algorithms scales as the number of elements \(n\) increases. We test both sequential and parallel versions on the Normal dataset with inputs ranging from 10k to 1B elements.

\begin{figure}[ht!]
\[
\resizebox{\textwidth}{!}{\input{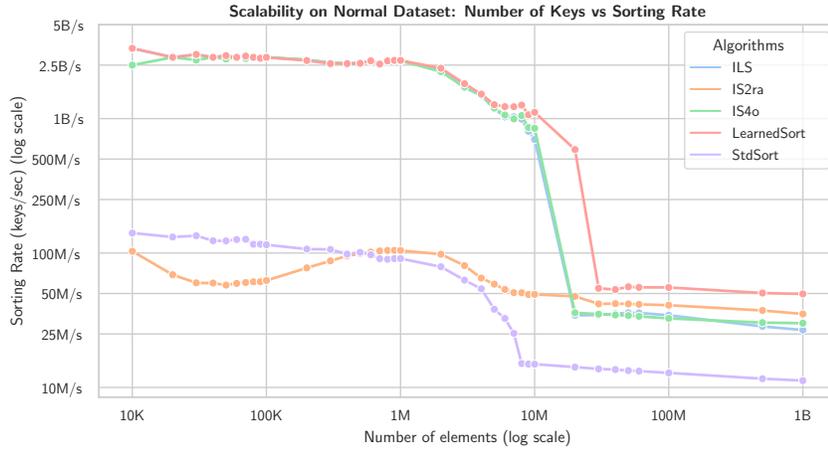}}
\]
\caption{Comparison of the sorting rates for the Normal dataset as the input size grows under a sequential setting. Higher rates are better.}
\end{figure}

On the sequential setting, Learned Sort, \(\text{IS}^{4}\text{o}\), and ILS have close sorting rates until the input hits a size of 10M elements. The mark of 10M elements is the point where the number of elements stops fitting into the cache. All algorithms except \(\text{IS}^{2}\text{Ra}\) have their sorting rates significantly reduced at that point. Learned Sort takes the lead at that point and \(\text{IS}^{2}\text{Ra}\) climbs to second. \(\text{IS}^{4}\text{o}\) and ILS remain tied albeit the former performs slightly faster at 1B elements.

Overall, ILS scales well in the sequential setting. Its sorting rate is almost identical to \(\text{IS}^{4}\text{o}\), which is very competitive because \(\text{IS}^{4}\text{o}\) is a state-of-the-art sorting algorithm. 

On the parallel setting, the sorting rates at the beginning are impacted by the overhead of using multiple threads. We note that \(\text{IPS}^{4}\text{o}\) and IPLS are slower than their sequential implementations at 1M elements, which puts the parallel implementation of \texttt{std::sort} at the top. However, we point out that the single-threaded versions of \(\text{IPS}^{4}\text{o}\) and IPLS are 2.5x faster than the parallel \texttt{std::sort} for the range of 1M to 10M elements.

\begin{figure}[H]
\[
\resizebox{\textwidth}{!}{\input{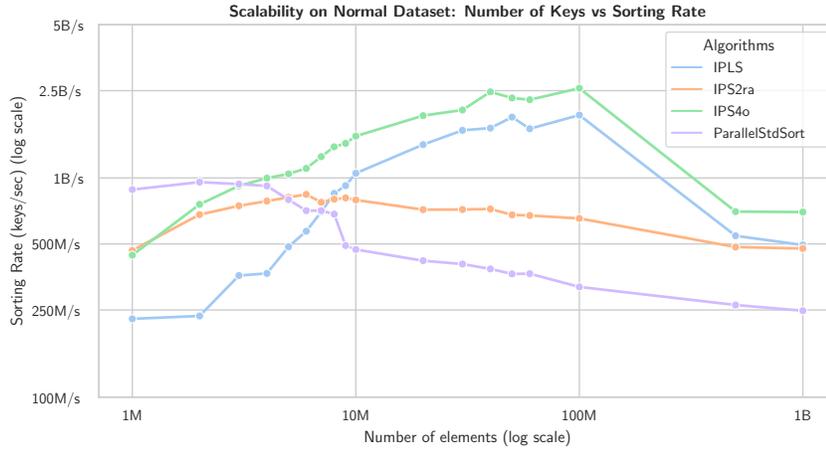}}
\]
\caption{Comparison of the sorting rates for the Normal dataset as the input size grows under a parallel setting. Higher rates are better.}
\end{figure}

As the input size grows, so does the sorting rate of IPLS and \(\text{IPS}^{4}\text{o}\) until they hit 100M elements. We interpret the growth as a sign that larger inputs make it simpler for the algorithms to utilize all the hardware resources available. The bottleneck during this range is not the partitioning but the scheduling of the parallel tasks.

After 100M elements are reached, the sorting rate starts to degrade. The cache becomes the limiting factor one more time and the gap between IPLS and \(\text{IPS}^{2}\text{Ra}\) shrinks, but IPLS always stays ahead of \(\text{IPS}^{2}\text{Ra}\). 

Overall, IPLS also scales well in the parallel setting. Its sorting rate in the 1M to 10M could be improved by limiting the number of threads being used, as the overhead of using all threads makes the algorithm slower than in the sequential setting. For the 10M to 1B range, IPLS does not win against \(\text{IPS}^{4}\text{o}\) but still beats \(\text{IPS}^{2}\text{Ra}\) and parallel \texttt{std::sort}.

\chapter{Conclusion and Future Work}

We have shown that the \(\text{IPS}^{4}\text{o}\) provides a framework to implement parallel
ML-enhanced sorting. Through the framework, we were able to achieve parallel learned sorting overcoming the limitations of earlier work that could not run in parallel.

The choice of linear models trained with FMCD did not crown our proposed algorithm as the fastest sorting algorithm available. However, it remained competitive against its peers. Remarkably, using a simple linear model to partition the data can achieve such sorting rates.

In this sense, our approach is promising because it lets us focus on improving the models used to sort and let the \(\text{IPS}^{4}\text{o}\) framework handle implementation details such as in-place partitioning and parallelism.

Future work to improve ML-enhanced sorting is an invitation to experiment with modeling. There is a gap in modeling the eCDF distribution of strings, as current models focus on numeric types. Moreover, there is room to incorporate ideas and models from the field of Learned Indexes into ML-enhanced sorting. Progress on Learned Indexes will benefit ML-enhanced sorting and vice-versa.

%

\newpage 
\pagestyle{fancy}\chead{Bibliography}\rhead{}\cfoot{}\rfoot{\thepage}

\bibliographystyle{ubco}
\bibliography{bibliography}

\newpage
\pagestyle{headings}
\addtocontents{toc}{%
\protect\renewcommand*\protect\cftchappresnum{\appendixname~}}

\addtocontents{toc}{
\setlength{\cftbeforechapskip}{\cftbeforesecskip}
\setlength{\cftchapindent}{\cftsecindent}
\protect\renewcommand{\cftchapfont}{\cftsecfont}
\protect\renewcommand{\protect\cftchapdotsep}{\cftsecdotsep}
}

\end{document}